\documentclass[11pt]{article}

\usepackage[margin=1in]{geometry}
\usepackage{microtype}

\usepackage{amsmath,amssymb}
\usepackage{graphicx}
\usepackage{booktabs}
\usepackage{multirow}

\usepackage{url}
\usepackage[hidelinks]{hyperref}

\newcommand{\etal}{et al.}

\newcommand{\keywordsline}[1]{\vspace{0.5em}\noindent\textbf{Keywords:} #1}

\title{\textbf{Bengali-Loop: Community Benchmarks for Long-Form Bangla ASR and Speaker Diarization}}

\author{%
\parbox{\textwidth}{\centering\scriptsize
H.M.~Shadman Tabib\textsuperscript{$*$},\;
Istiak Ahmmed Rifti\textsuperscript{$*$},\;
Abdullah Muhammed Amimul Ehsan\textsuperscript{$*$},\;
Somik Dasgupta,\;
Md~Zim Mim Siddiqee Sowdha\textsuperscript{$\dagger$},\;
Abrar Jahin Sarker\textsuperscript{$\dagger$},\;
Md.~Rafiul Islam Nijamy\textsuperscript{$\dagger$},\;
Tanvir Hossain\textsuperscript{$\dagger$},\;
Mst.~Metaly Khatun\textsuperscript{$\dagger$},\;
Munzer Mahmood\textsuperscript{$\dagger$},\;
Rakesh Debnath\textsuperscript{$\dagger$},\;
Gourab Biswas\textsuperscript{$\dagger$},\;
Asif Karim\textsuperscript{$\dagger$},\;
Wahid Al Azad Navid\textsuperscript{$\dagger$},\;
Masnoon Muztahid\textsuperscript{$\dagger$},\;
Fuad Ahmed Udoy\textsuperscript{$\dagger$},\;
Shahad Shahriar Rahman\textsuperscript{$\dagger$},\;
Md.~Tashdiqur Rahman Shifat\textsuperscript{$\dagger$},\;
Most.~Sonia Khatun\textsuperscript{$\dagger$},\;
Mushfiqur Rahman\textsuperscript{$\dagger$},\;
Md.~Miraj Hasan\textsuperscript{$\dagger$},\;
Anik Saha\textsuperscript{$\dagger$},\;
Mohammad Ninad Mahmud Nobo\textsuperscript{$\dagger$},\;
Soumik Bhattacharjee\textsuperscript{$\dagger$},\;
Tusher Bhomik\textsuperscript{$\dagger$},\;
Ahmmad Nur Swapnil\textsuperscript{$\dagger$},\;
Shahriar Kabir\textsuperscript{$\dagger$}\\[0.6em]
\scriptsize Department of Computer Science and Engineering, Bangladesh University of Engineering and Technology\\[0.3em]
\scriptsize\textsuperscript{$*$}Equal first-author contribution.\quad \textsuperscript{$\dagger$}Equal contribution (from Md~Zim Mim Siddiqee Sowdha onwards).%
}%
}

\date{} 

\begin{document}
\maketitle

\begin{abstract}
Bengali (Bangla) remains under-resourced in long-form speech technology despite its wide use.
We present Bengali-Loop, two community benchmarks to address this gap: (1) a \textbf{long-form ASR} corpus of 191 recordings (158.6 hours, 792k words) from 11 YouTube channels, collected via a reproducible subtitle-extraction pipeline and human-in-the-loop transcript verification; and (2) a \textbf{speaker diarization} corpus of 24 recordings (22 hours, 5{,}744 annotated segments) with fully manual speaker-turn labels in CSV format.
Both benchmarks target realistic multi-speaker, long-duration content (e.g., Bangla drama/natok).
We establish baselines (Tugstugi: 34.07\% WER; pyannote.audio: 40.08\% DER) and provide standardized evaluation protocols (WER/CER, DER), annotation rules, and data formats to support reproducible benchmarking and future model development for Bangla long-form ASR and diarization.
\end{abstract}

\keywordsline{Bengali, Bangla, long-form ASR, speaker diarization, benchmark, low-resource speech}

\section{Introduction}
Bengali (Bangla) is among the most spoken languages worldwide \cite{bengali_wikipedia} but remains under-resourced in open speech technology \cite{besacier2014underresourced}, especially for long-form automatic speech recognition (ASR) and speaker diarization.
Most public Bengali corpora focus on short, single-speaker utterances (often read speech) from crowdsourcing \cite{openslr53,alam2022commonvoice}.
While useful for conventional ASR, such data does not capture the challenges of long recordings: multiple topics, varied acoustics, multi-speaker dialogue, and error accumulation over time.
Speaker diarization (``who spoke when'') is crucial for archives, interviews, meetings, and broadcast content, yet Bengali diarization resources remain scarce compared with benchmarks like AMI \cite{carletta2005ami} and DIHARD \cite{ryant2021dihard3}.

Recent multilingual ASR models have improved recognition across many languages \cite{radford2022whisper,baevski2020wav2vec,babu2022xlsr}, and community efforts have released large Bengali ASR datasets \cite{ardila2019commonvoice,openslr53,alam2022commonvoice,rakib2023oodspeech}.
However, these resources focus mainly on ASR; fully manual speaker diarization labels are rare for Bengali, and long-form (tens of minutes to hours) continuous-transcript settings remain underexplored.

\textbf{Contributions.} We introduce Bengali-Loop, two community benchmarks:
\begin{itemize}
  \item \textbf{Long-form ASR:} 191 recordings (158.6 hours, 792k words, 36k vocabulary) from 11 YouTube channels, collected via a reproducible subtitle-extraction pipeline and human-in-the-loop transcript curation.
  \item \textbf{Speaker diarization:} 24 recordings (22 hours total, 5{,}744 segments) with fully manual annotations---10 train (9.5 hours) and 14 test (12.5 hours)---in CSV format (start time, end time, speaker id), averaging 16 speakers per recording.
  \item Standardized formats, annotation rules (including overlap policy), and evaluation protocols (WER/CER, DER) for reproducible benchmarking.
  \item Baseline results: Tugstugi and Hishab TITU-BN for ASR; pyannote.audio and two lighter pipelines (WebRTC/Silero VAD + ECAPA clustering) for diarization \cite{tugstugi_model,bredin2023pyannote}.
\end{itemize}

\section{Related Work}
\subsection{Bengali ASR corpora}
Bengali speech resources have grown through crowdsourcing and community efforts.
OpenSLR SLR53 offers a large training set with about 196k utterances \cite{openslr53}; Common Voice \cite{ardila2019commonvoice} and its Bengali subset \cite{alam2022commonvoice} provide crowdsourced data; OOD-Speech \cite{rakib2023oodspeech} offers an out-of-distribution benchmark with 1178 hours training and 23-hour test from diverse media.
These corpora typically focus on short utterances or segmented clips \cite{panayotov2015librispeech}; long-form continuous-transcript evaluation remains limited.

\subsection{Speaker diarization}
Diarization benchmarks such as AMI \cite{carletta2005ami}, VoxCeleb \cite{nagrani2017voxceleb}, VoxConverse \cite{chung2020voxconverse}, and DIHARD \cite{ryant2021dihard3} have advanced the field but do not target Bengali.
Bengali diarization resources with gold speaker-turn annotations remain scarce.
We introduce a fully manual Bengali diarization benchmark suitable for DER evaluation and reproducible baselines.

\subsection{Toolkits}
Whisper \cite{radford2022whisper} and XLS-R \cite{babu2022xlsr} (building on wav2vec~2.0 \cite{baevski2020wav2vec}) provide strong multilingual baselines for Bengali.
For diarization, pyannote.audio \cite{bredin2023pyannote} offers widely used pretrained pipelines; we also use ECAPA-TDNN embeddings \cite{desplanques2020ecapa} via SpeechBrain \cite{ravanelli2021speechbrain} for lighter baselines.

\section{Bengali-Loop Long-Form ASR Dataset}
\subsection{Domains and sources}
The long-form ASR corpus is drawn from public Bengali media, primarily Bangla drama (natok), with contributions from news, short films, and entertainment channels.
This introduces real-world variability: conversational speech, background music, studio recordings, and multi-speaker dialogue.

\subsection{Collection and processing pipeline}
We collect long-form audio-video content with Bengali subtitles (creator-provided or auto-generated), extracting the audio track and subtitle tracks (e.g., WebVTT).
Subtitles are parsed into raw transcripts by stripping timestamps, removing formatting cues (e.g., music markers), resolving HTML entities, and merging duplicated caption lines.
Audio is standardized to 16\,kHz mono WAV.
Each item is logged in a structured manifest (IDs, durations, source fields, paths) for reproducibility.

\subsection{Human-in-the-loop transcript curation}
Raw subtitles are insufficient as ground-truth due to omissions, spelling variation, and noise.
We perform offline human verification and correction under the following guidelines:
\begin{itemize}
  \item \textbf{Correct subtitle errors} to match the spoken audio.
  \item \textbf{Add missing words/phrases} if the subtitle omits content.
  \item \textbf{Numbers:} convert digits into Bengali word forms for consistent scoring.
\end{itemize}
This yields higher-quality reference transcripts at lower cost than transcribing from scratch.

\subsection{Data format}
Each recording has an audio file and a transcript file. We also provide a JSONL manifest:
\begin{verbatim}
{
  "id": "example_0001",
  "audio_filepath": "audio/example_0001.wav",
  "duration": 1784.5,
  "text": "TRANSCRIPT_TEXT"
}
\end{verbatim}

\subsection{Dataset statistics}
The raw corpus comprises 191 long-form recordings totalling 158.6 hours from 11 YouTube channels.
Table~\ref{tab:asr_raw_stats} gives the key statistics.

\begin{table}[t]
  \caption{Bengali-Loop long-form ASR raw dataset summary.}
  \label{tab:asr_raw_stats}
  \centering
  \begin{tabular}{lr}
    \toprule
    \textbf{Statistic} & \textbf{Value} \\
    \midrule
    Total recordings          & 191 \\
    Total duration (hours)    & 158.6 \\
    Mean duration (min)       & 49.8 \\
    Median duration (min)     & 43.4 \\
    Min duration (min)        & 1.2 \\
    Max duration (min)        & 306.8 \\
    Total words               & 792{,}491 \\
    Total characters          & 4{,}089{,}938 \\
    Vocabulary size            & 36{,}050 \\
    Mean words / recording    & 4{,}149 \\
    Unique source channels    & 11 \\
    \bottomrule
  \end{tabular}
\end{table}

\begin{table}[t]
  \caption{Source channel distribution in the Bengali-Loop raw corpus.}
  \label{tab:channel_dist}
  \centering
  \begin{tabular}{lrc}
    \toprule
    \textbf{Channel} & \textbf{\#Recs} & \textbf{\%} \\
    \midrule
    Eagle Premier Station & 61 & 31.9\% \\
    Banglavision DRAMA & 47 & 24.6\% \\
    Maasranga Drama & 31 & 16.2\% \\
    CMV & 19 & 9.9\% \\
    KS Entertainment & 12 & 6.3\% \\
    GOLLACHUT & 9 & 4.7\% \\
    Raad Drama & 6 & 3.1\% \\
    Rabbit Entertainment & 3 & 1.6\% \\
    Others (3 channels) & 3 & 1.6\% \\
    \bottomrule
  \end{tabular}
\end{table}

\begin{table}[t]
  \caption{Distribution of recording durations.}
  \label{tab:dur_buckets}
  \centering
  \begin{tabular}{lrc}
    \toprule
    \textbf{Duration Range} & \textbf{\#Recs} & \textbf{\%} \\
    \midrule
    0--20 min & 10 & 5.2\% \\
    20--40 min & 47 & 24.6\% \\
    40--60 min & 104 & 54.5\% \\
    60--80 min & 18 & 9.4\% \\
    80--100 min & 5 & 2.6\% \\
    $>$100 min & 7 & 3.7\% \\
    \bottomrule
  \end{tabular}
\end{table}

\begin{figure}[t]
  \centering
  \includegraphics[width=\textwidth]{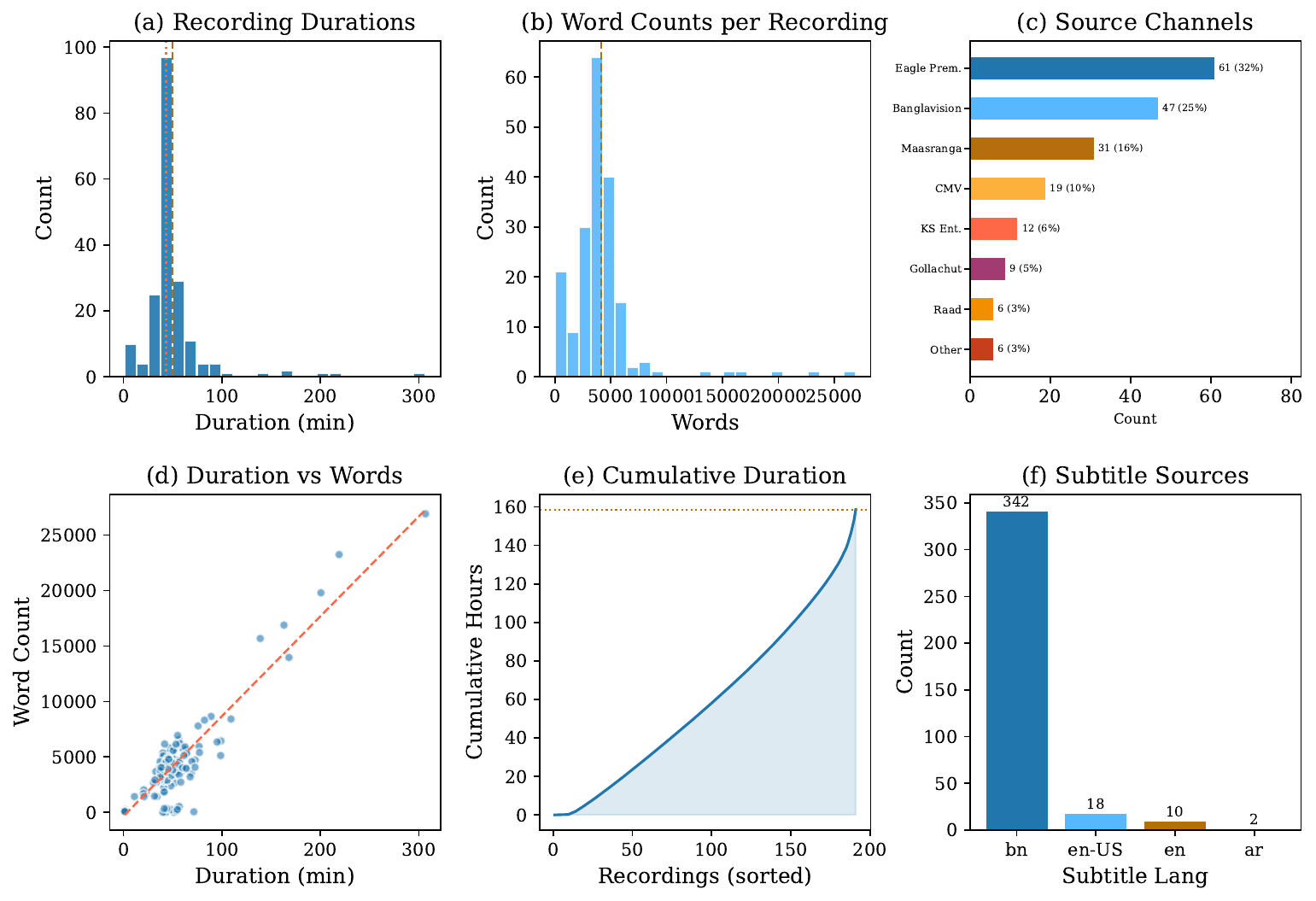}
  \caption{Bengali-Loop raw dataset statistics. (a) Distribution of recording durations (mean and median marked). (b) Distribution of transcript word counts per recording. (c) Source channel breakdown (abbreviations: Eagle Prem.\ = Eagle Premier Station; Banglavision = Banglavision DRAMA; Maasranga = Maasranga Drama; KS Ent.\ = KS Entertainment; Gollachut = GOLLACHUT; Raad = Raad Drama; Rabbit Ent.\ = Rabbit Entertainment; Club 11 = Club 11 Entertainment; Folk Studio = Folk Studio Bangla). (d) Duration vs.\ word count with linear fit. (e) Cumulative duration across recordings. (f) Subtitle source language distribution.}
  \label{fig:dataset_stats}
\end{figure}

The majority of recordings (54.5\%) fall in the 40--60 minute range (Table~\ref{tab:dur_buckets}), consistent with typical Bangla drama (natok) lengths.
Figure~\ref{fig:dataset_stats} summarises the corpus: (a)--(b) duration and word-count distributions, (c) source channel breakdown, (d) duration--word relationship, (e) cumulative hours, and (f) subtitle source languages.

\section{Bengali-Loop Speaker Diarization Dataset}
\subsection{Task definition}
Given a multi-speaker long-form Bangla recording, the task is to produce speaker-homogeneous segments specifying ``who spoke when''.
Speaker identities are anonymous and recording-local: IDs are consistent only within each recording.

\subsection{Fully manual annotation protocol}
All diarization annotations are produced end-to-end by human annotators:
\begin{itemize}
  \item Annotators mark uninterrupted regions where a single speaker speaks without interruption.
  \item A new segment starts when a different speaker begins speaking.
  \item Speaker IDs start at 1. New speakers get new IDs, and returning speakers reuse their prior ID within the same recording.
  \item \textbf{Overlap policy:} when multiple speakers overlap simultaneously, we assign the segment to the \emph{first speaker} who started speaking for that overlapped period (single-label reference).
\end{itemize}

\subsection{Annotation format}
Each recording has an accompanying CSV with columns \texttt{start\_time}, \texttt{end\_time} (HH:MM:SS), and \texttt{speaker\_id} (integer).

\subsection{Dataset statistics}
The diarization corpus has 10 train recordings (9.5 hours, 2{,}612 segments, 10--22 speakers each) and 14 test recordings (12.5 hours, 3{,}132 segments, 1--27 speakers each).
Table~\ref{tab:diar_stats} summarises the splits.

\begin{table}[t]
  \caption{Bengali-Loop diarization dataset statistics.}
  \label{tab:diar_stats}
  \centering
  \begin{tabular}{lcccc}
    \toprule
    \textbf{Split} & \textbf{\#Recs} & \textbf{Hours} & \textbf{Segments} & \textbf{Avg. spkrs/rec} \\
    \midrule
    Train & 10 & 9.5 & 2{,}612 & 17.1 \\
    Test  & 14 & 12.5 & 3{,}132 & 15.4 \\
    \midrule
    Total & 24 & 22.0 & 5{,}744 & 16.1 \\
    \bottomrule
  \end{tabular}
\end{table}

\section{Benchmarking and Evaluation Protocol}
\subsection{Long-form ASR evaluation (WER/CER)}
We report word error rate (WER) and character error rate (CER) \cite{panayotov2015librispeech} with explicit normalization: Unicode (NFC), punctuation removal, consistent whitespace, digits-to-words, and code-mixed token handling.
WER is defined as:
\begin{equation}
\mathrm{WER} = \frac{S + D + I}{N} \times 100\%,
\end{equation}
where $S$ is substitutions, $D$ deletions, $I$ insertions, and $N$ the number of reference words.

\textbf{Long-form decoding.} For models with limited context (e.g., Whisper), we use chunked inference (e.g., 30-second overlapping windows) and merge hypotheses with overlap-aware stitching.
An official scoring script will be released for consistent reporting.

\subsection{Speaker diarization evaluation (DER)}
We report diarization error rate (DER) \cite{fiscus2006nist} with the following settings:
\begin{itemize}
  \item a collar (e.g., 0.25 s) at segment boundaries,
  \item optimal speaker mapping (Hungarian) between system and reference speaker IDs,
  \item single-label scoring aligned with our overlap policy (overlaps attributed to the first speaker).
\end{itemize}
FA/MISS/CONF breakdowns may be reported for error analysis.

\section{Baseline Systems}
\subsection{ASR baselines}
We evaluate two Bengali ASR systems on 37 test recordings: Tugstugi (dialect-trained Whisper) \cite{tugstugi_model} and Hishab TITU-BN.
Results are given in Table~\ref{tab:asr_results}.

\begin{table}[t]
  \caption{ASR baseline results on Bengali-Loop test set.}
  \label{tab:asr_results}
  \centering
  \begin{tabular}{lcc}
    \toprule
    \textbf{Model} & \textbf{WER (\%)} & \textbf{CER (\%)} \\
    \midrule
    Tugstugi & 34.07 & 16.44 \\
    Hishab TITU-BN & 50.67 & 21.99 \\
    \bottomrule
  \end{tabular}
\end{table}

\subsection{Diarization baselines}
\textit{VAD: Voice Activity Detection; ECAPA: Emphasized Channel Attention, Propagation and Aggregation; DER: Diarization Error Rate.}
\medskip
We evaluate three diarization systems on the test set:
\begin{itemize}
  \item \textbf{pyannote.audio} (pretrained pipeline) \cite{bredin2023pyannote}.
  \item  Silero VAD $\rightarrow$ ECAPA embeddings $\rightarrow$ agglomerative clustering.
  \item  WebRTC VAD $\rightarrow$ SpeechBrain ECAPA embeddings $\rightarrow$ agglomerative clustering.
\end{itemize}

Table~\ref{tab:diar_results} reports DER on the 14-recording test set.
pyannote.audio achieves the lowest DER (40.08\%), while the lighter pipelines reach 61.50\% and 73.71\%, highlighting both the benchmark difficulty and the potential for Bengali-specific diarization models.

\begin{table}[t]
  \caption{Diarization baseline results on Bengali-Loop test set.}
  \label{tab:diar_results}
  \centering
  \begin{tabular}{lc}
    \toprule
    \textbf{System} & \textbf{DER (\%)} \\
    \midrule
    pyannote.audio (pretrained) & 40.08 \\
    WebRTC VAD + ECAPA + Clustering &  73.71 \\
    Silero VAD + ECAPA + Clustering & 61.50 \\
    \bottomrule
  \end{tabular}
\end{table}

\section{Release, Licensing, and Reproducibility}
\textbf{Data availability.} The benchmarks were released as part of an organized competition (DL Sprint~4.0). The ASR corpus is available at \href{https://www.kaggle.com/competitions/dl-sprint-4-0-bengali-long-form-speech-recognition/data}{DL Sprint~4.0 Bengali Long-Form Speech Recognition}; the diarization corpus at \href{https://www.kaggle.com/competitions/dl-sprint-4-0-bengali-speaker-diarization-challenge/data}{DL Sprint~4.0 Bengali Speaker Diarization}.

\textbf{Source media and redistribution.} Audio is drawn from publicly accessible Bengali media. Where redistribution is constrained, we release annotations, transcripts, manifests, and reproducible scripts rather than raw media.

\textbf{Reproducibility.} We plan to provide scripts to reproduce the collection manifest, standardize audio to 16\,kHz mono WAV, apply text normalization for scoring, and run baseline inference and evaluation.

\textbf{Takedown.} A takedown channel is documented so that rights holders may request removal of identifiers or annotations associated with their content.

\section{Ethics and Responsible Use}
We avoid sensitive personal data and encourage responsible development.
Documentation covers data composition, intended use, limitations, and the takedown mechanism.
We ask benchmark users to respect original source licenses and to report results using the official evaluation settings for comparability.

\section{Limitations}
The release has limited dialectal diversity.
The diarization reference uses a single-label overlap policy (attributing overlap to the first speaker), which simplifies annotation and scoring but does not label multi-speaker overlap explicitly.
Domain balance (drama, news, short films) may not reflect all Bengali use cases, and long-form decoding choices can affect WER unless standardized.

\section{Conclusion}
We presented Bengali-Loop, two community benchmarks for Bangla speech technology: (1) a long-form ASR corpus of 191 recordings (158.6 hours, 792k words) with human-verified transcripts, and (2) a fully manual diarization corpus of 24 recordings (22 hours, 5{,}744 segments) with speaker-turn annotations.
Baselines establish 34.07\% WER (Tugstugi) and 40.08\% DER (pyannote.audio) as initial performance anchors.
Standardized formats, protocols, and evaluation scripts support reproducible benchmarking and future development for Bangla long-form ASR and diarization.

\section*{Acknowledgments}
We thank the annotators and volunteers who contributed to transcript verification and diarization labeling.
We acknowledgeSomik Dasgupta, Swarup Sidhartho Mondol, Md Zim Mim Siddiqee Sowdha, Wahid Al Azad Navid, Ahmmad Nur Swapnil, Gourab Biswas, Ananya Shahrin Promi, Jarin Tasneem, H.M. Shadman Tabib, and Jarin Tasnim for performing voiceovers over scripts to build the hidden test cases (Link will be available later)


\end{document}